\documentclass[11pt,a4paper]{article}
\pdfoutput=1
\usepackage{jheppub,hyperref}
\usepackage{multirow, graphicx,amssymb,url,mathrsfs,amsmath}
\usepackage{wrapfig,boxedminipage,setspace,subfigure,epsfig}
\usepackage{amsxtra,amstext,latexsym,dsfont,amsfonts}
\usepackage{color,eucal}
\usepackage[section]{placeins}
\usepackage[colorlinks=true,linktocpage=true,linkcolor=blue,citecolor=blue]{hyperref}
\graphicspath{{Figures/}}
\newcommand{\mt}[1]{\textrm{\tiny #1}}
\newcommand{\be}{\begin{equation}}
\newcommand{\ee}{\end{equation}}
\newcommand{\bea}{\begin{eqnarray}}
\newcommand{\eea}{\end{eqnarray}}
\newcommand{\rh}{r_\mt{H}}

\title{The magnetic effect on strongly correlated system with two currents near the QCP from holography } 
\author[a]{Jaeha Lee,}
\author[b]{Sang-Jin Sin,}
\author[b]{Geunho Song}
\emailAdd{jaeha@caltech.edu}
\emailAdd{sjsin@hanyang.ac.kr}
\emailAdd{sgh8774@gmail.com}
\affiliation[a]{The division of physics, mathematics and astronomy, California Institute of Technology, Pasadena, CA 91125, USA}
\affiliation[b]{ Department of Physics, Hanyang University, Seoul 04763, South Korea }
 \abstract{   
We study the magnetic effect in strongly interacting  system with two conserved currents near the quantum critical point (QCP). For this purpose, we introduce  the hyper-scaling violation geometry with the blackhole. Considering the perturbation near the background geometry, we compute the transport coefficients using holographic methods.
We calculated the magneto-transport for general QCP and discuss the special point $(z,\theta)=(3/2,1)$ where the data of Dirac material were well-described previously.
}
\keywords{Gauge/Gravity duality, hyper-scaling violation, quantum critical points, transports}
\begin{document}
\maketitle
 
\section{Introduction} 
 
For the strongly correlated system,  the particle nature often  is absent so that   theories based on quasi-particles such as Landau-Fermi liquid theory fail. The strong correlation can happen even for weakly interacting system when the Fermi surface can be tuned to be very small, because then the electron-hole pairs which  screen the Coulomb interaction are not sufficiently created due to the smallness of the Fermi surface.  Therefore, any Dirac fluid can be strongly correlated as far as it has the small Fermi surface, which is already shown in the clean graphene~\cite{Lucas:2015sya,pkim} and in the surface of the topological insulator with magnetic doping ~\cite{liu2012crossover,zhang2012interplay,bao2013quantum}. 
  To describe such  system, we need to find a new way. We  consider the quamtum critical point (QCP), where the microscopic details in UV are  irrelevant and most of the information in UV is apprently  lost in low energy probe in the sense of the coarse graining. This apparent loss of information is very similar to black hole system and this similarity between a QCP and a blackhole is the important motivation to use the holography to analyze the strongly correlated electron system. A QCP can be characterized by $(z,\theta)$ which is defined by the dispersion relation $\omega \sim k^z$ and the entropy density $s\sim T^{(d-\theta)/z}$. We can use a geometry with the same scaling symmetry respected $(t,r,x)\rightarrow (\lambda^z t, \lambda^{-1}r, \lambda x)$:
  \bea
  	ds^2=r^{-\theta}\left(-r^{2z}dt^2+\frac{dr^2}{r^2}+r^2 d\vec{x}^2\right).
  \eea
which is called hyper-scaling violation (HSV) geometry.

In our previous works~\cite{Seo:2016vks,Song2020,Seo:2017oyh,Ge2020}, we described the clean graphene and the topological insulator with magnetic doping in some parameter regions by using holographic method. For the surface of the topological insulator, we introduced just one current   with a interaction  to encode the magnetic doping ~\cite{Seo:2017oyh,Ge2020}. We calculated magneto-conductivity and investigated the phase transitions from weak localization to weak anti-localization. As in the case of the graphene, it turns out that we can have the better fit for $(z,\theta)=(3/2,1)$ than $(1,0)$.

For the graphene, we  need two currents model~\cite{Seo:2016vks,Song2020}: when the electron and hole densities fluctuate from their equilibrium states, the system is supposed to reduce the difference by creating or absorbing electron-pair:
\bea
	e^-\leftrightarrow e^-+h^++e^-, \qquad h^+\leftrightarrow h^++h^++e^-.
\eea 
In this process, the energy and momentum should be conserved. For the graphene, however, the kinematically available states are severely reduced  \cite{Foster} due to the geometry of the Dirac cone and this constraints makes the two currents $J_e$ and $J_h$    independently conserved. Hence, we need the two independent currents to describe the graphene.
In ~\cite{Song2020}, we analyzed the two currents model for hyperscaling violation geometry (HSV), and we found that the theory with a QCP at $(z,\theta)=(3/2,1)$ than $(1,0)$ studied in ~\cite{Seo:2017oyh}. 
   The value $\theta=1$ is important because  the holographic background with dual Fermi surface has the effective dimension $d_{eff}=d-\theta$ and the system with fermi surface should have $ d_{eff}=1$ so that $\theta=1$ can describe  the character of fermion in this aspect.    Indeed, we found that $(z,\theta)=(3/2,1)$ can fit the data better therefore qualified as   more proper critical exponent   both in graphene and topological insulator. 

In this paper, we study the holographic model with the two currents and a particular interactiion which is shown to describe  magnetically doped material\cite{Seo:2017oyh,Ge2020}. 
We calculated all transport coefficients and demonstrated some typical behavior of the magneto-transports. Although we don't have any experimental results for magnetically doped graphene, our work  can be considered as predictions for the magnetic effect for graphene or other material which need two or more currents due to the presence of two layers or two valleys. 

 \section {\bf The two currents model with magnetic impurity in hyperscaling violating geometry }  
 We   start from a 4-dimensional action with asymptotically hyperscaling geometry $g_{\mu\nu}$, which includes a dilaton field $\phi$,  a  gauge fields $A_{\mu}$ to complete the  asymptotic hyperscaling violating geometry, two extra gauge fields $B^{(a)}_{\mu}$ which are dual to two conserved currents, and the axion fields $\chi_1,$ $\chi_2$ to break the translational symmetry. 
\bea\label{action1}
S&=&\int_{\mathcal{M}}d^4x(\mathcal{L}_0+\mathcal{L}_{int})\nonumber\\
\mathcal{L}_0&=&\sqrt{-g}\left(R+\sum_{i=1}^2 V_ie^{\gamma_i\phi}-\frac{1}{2}(\partial\phi)^2-\frac{1}{4}Z_A F^2-\sum_a^2\frac{1}{4}Z_a G_{(a)}^2-\frac{1}{2}Y\sum\limits_{i}^{2}(\partial \chi_i)^2\right)\nonumber\\
\mathcal{L}_{int}&=&-\sum_{a, i=1,2}\frac{q_{\chi_a}}{16}(\partial\chi_i)^2\epsilon^{\mu\nu\rho\sigma}G^{(a)}_{\mu\nu}G^{(a)}_{\rho\sigma}
\eea
where $F=dA$, $G_{(a)}=dB_a$. 
We use the ansatz  
\be
Z_A= e^{\lambda \phi}, \quad Z_a=\bar{Z}_ae^{\eta\phi}, \quad Y=e^{-\eta\phi},\quad \chi_i=\beta x_i,
\ee
where $\beta$ denotes the strength of momentum relaxation.
The equations of motion for gauge fields and gravity  are given by
 \begin{eqnarray}
&&\partial_{\mu}(\sqrt{-g}g^{\mu\nu}Y\sum_{i}\partial_{\nu}\chi_i)+\sum_{a,i=1,2}\frac{q_{\chi_a}}{8}\partial_{\mu}(\epsilon^{\rho\sigma\lambda\gamma}G^{(a)}_{\rho\sigma}G^{(a)}_{\lambda\gamma}g^{\mu\nu}\partial_{\nu}\chi_i)=0,\\
&&\partial_{\mu}(\sqrt{-g}Z_AF^{\mu\nu})=0, \quad 
\partial_{\mu}(\sqrt{-g}Z_aG^{\mu\nu}_{(a)}+\frac{q_{\chi_a}}{4}\sum_i(\partial\chi_i)^2\epsilon^{\alpha\beta\mu\nu}G^{(a)}_{\alpha\beta})=0,\label{max}  \\
&&R_{\mu\nu}=\frac{1}{2\sqrt{-g}}g_{\mu\nu}\mathcal{L}_0+\frac{1}{2}\partial_{\mu}\phi\partial_{\nu}\phi+\frac{Y}{2}\sum_i \partial_{\mu}\chi_i\partial_{\nu}\chi_i+\frac{1}{2}Z_A F_{\mu}^{\rho}F_{\nu \rho}+\sum_a^2\frac{1}{2}Z_a G_{(a)\mu}^{\rho}G_{\nu \rho}^{(a)}\nonumber\\
&&+\sum_{a,i=1,2}\frac{q_{\chi_a}}{16}\frac{1}{\sqrt{-g}}(\partial_{\mu}\chi_i)(\partial_{\nu}\chi_i)\epsilon^{\rho\sigma\lambda\gamma}G^{(a)}_{\rho\sigma}G^{(a)}_{\lambda\gamma}\\
&& \Box\phi+\sum_i V_i\gamma e^{\gamma\phi}-\frac{1}{4} Z_A'(\phi) F^2-\frac{1}{4}\sum_a Z_a'(\phi) G^2_{(a)}-\frac{1}{2}Y'(\phi)\sum\limits_{i}^{2}(\partial \chi_i)^2=0.
\end{eqnarray}
The solution for the dilaton field is given by 
\be
\phi(r)=  \nu\ln r, \quad \hbox{with } \nu=\sqrt{(2-\theta)(2z-2-\theta)}.
\ee
By solving the equations of motion, we can get the gauge couplings and dilaton coupling $Z_A$, $Z_a$, and $Y$ as followings:
\bea
Z_A(\phi)=e^{\lambda\phi}= r^{\theta-4},\quad 
Z_a(\phi)=\bar{Z}_ae^{\eta\phi}=\bar{Z}_ar^{2z-\theta-2}, \quad 
Y(\phi)=e^{-\eta\phi},
\eea 
where $\lambda=(\theta-4)/\nu$, $\eta=\nu /(2-\theta)$.

Other exponents and potentials are given by
\bea 
\gamma_1=\frac{\theta}{\nu }, \quad \gamma_2=\frac{\theta+2z-6}{\nu},\quad V_1=\frac{z-\theta+1}{2(z-1)}q_A^2, \quad V_2=\frac{H^2(2z-\theta-2)}{4(z-2)}
\eea
where $H$ is a constant magnetic field and $q_A=\sqrt{2(-1+z)(2+z-\theta)}$. 
Finally, we have the following background solutions:
\begin{eqnarray}\label{bgsol}
&&A=a(r)dt,  ~B_a=b_a(r)dt-\frac{1}{2}H y dx+\frac{1}{2}H x dy,   \\
&&\quad \chi=(\beta x, \beta y),  \\
&&ds^{2}=r^{-\theta}\bigg(-r^{2z}f(r)dt^{2}+\frac{dr^{2}}{r^2 f(r)}+r^{2}(dx^{2}+dy^{2})\bigg), \\
&&f(r)=1-m r^{\theta-z-2}-\frac{\beta^{2}}{(\theta-2)(z-2)}r^{\theta-2z}+\frac{(\bar{Z}_1q_1^{2}+\bar{Z}_2q_2^2)(\theta-z)r^{2\theta-2z-2}}{2(\theta-2)}\nonumber\\
&&+\frac{(\bar{Z}_1+\bar{Z}_2)H^2r^{2z-6}}{4(z-2)(3z-\theta-4)}-\frac{c_2\beta^2H(q_{\chi_1}q_1+q_{\chi_2}q_2)}{r^{4+2z-3\theta}}+\frac{c_3\beta^4H^2(\frac{q_{\chi_1}}{\bar{Z}_1}+\frac{q_{\chi_2}}{\bar{Z}_2})}{r^{6+2z-4\theta}}\nonumber\\
&&a(r)=\frac{-q_A}{2+z-\theta}(\rh^{2+z-\theta}-r^{2+z-\theta}), \;\; b_a(r)= \mu_a-q_a r^{\theta-z}-\frac{c_4 q_{\chi_a}\beta^2 H}{\bar{Z}_ar^{z-2\theta+2}} ,
\end{eqnarray}
where $a=1,2$ and $c_2$, $c_3$, $c_4$ are given by
\bea
	c_2 = \frac{(z-\theta)}{(\theta-2)(2\theta-z-2)},\quad c_3 = \frac{1}{2(2-\theta)(4+z-3\theta)},\quad c_4=\frac{1}{2\theta-z-2}
\eea

\begin{figure}
\begin{center}
\subfigure[The scheme of embedding HSV to AdS]{\includegraphics[angle=0,width=0.5\textwidth]{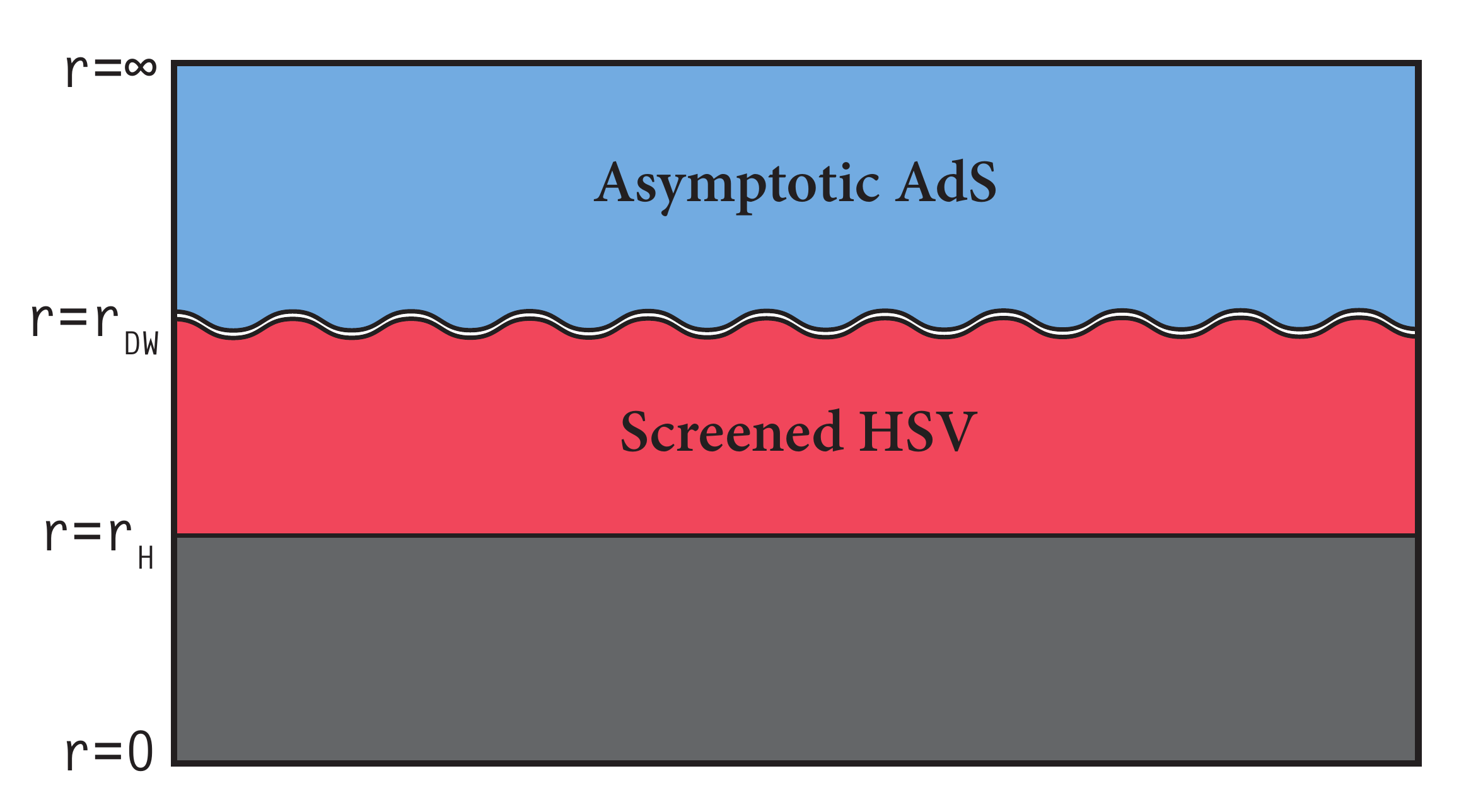}}
\caption{The schematic figure of our geometry. The region inside the blackhole is colored with gray and there is a domain wall $(r=r_{DW})$ somewhere between $AdS$ boundary and blackhole horizon at $r_{H}$. 
   } \label{region} 
\end{center}
\end{figure}

This HSV solution should be embedded into asymptotically AdS spacetime so that it is just the IR part of the total domain-wall solution. Here, we only conceptually embedded but did not write down the explicit solution in the entire region, since it is not important for the computation of the DC transports~\cite{Donos:2014cya}. 
See figure \ref{region}.

We can define the conserved charge from the equations of motion for the gauge fields $B_a$ as the constants of integration
\bea
	Q_a &&= \sqrt{-g}Z_aG^{tr}_{(a)}+\frac{q_{\chi_a}}{4}\sum_i(\partial \chi_i)^2\epsilon^{\alpha\beta t r}G_{\alpha\beta}^{(a)}\nonumber\\
	        &&=\bar{Z}_aq_a(z-\theta) = (z-\theta)\left(\mu_a \bar{Z}_a \rh^{z-\theta}+\frac{q_{\chi_a}\beta^2\rh^{-2+\theta}}{2+z-2\theta}\right)
\eea
The entropy density and the Hawking temperature are given by
\bea
s=&&4\pi r^{2-\theta}_H, \\
{4\pi}T=&& (z+2-\theta)\rh^z -\frac{\beta^2 \rh^{\theta-z}}{2-\theta}
-\frac{\rh^{2\theta-2-z}}{2(2-\theta)}\sum_{a=1,2}\frac{1}{\bar{Z}_a}\left(\Theta_a H -Q_a\right)^2 - \bar{Z}\frac{H^2\rh^{3z-6}}{4(2-z)}
\label{r0Trelation}
\eea
where $\Theta_a=q_{\chi_a}\beta^2\rh^{\theta-2}$ and $\bar{Z}=\sum_{a=1,2}\bar{Z_a}$.

\section{Conserved currents and DC transports}
We consider following perturbations to compute the transport coefficients based on the idea of linear response theory ~\cite{Donos:2014cya}:
\bea
	&&\delta g_{ti}=h_{ti}(r)+tf_{3i}(r),~~ \delta g_{ri}=h_{ri}(r), ~~ \delta B_{ai}=\tilde{b}_{ai}-tf_{ai},~~ \delta\chi_i=\varphi_{i}(r).
\eea
 We take the functions $f_i(r)$ as
\bea
f_{1i}&=&-E_{1i}+\zeta_i b_1(r)\nonumber\\
f_{2i}&=&-E_{2i}+\zeta_i b_2(r)\nonumber\\
 f_{3i}&=&-\zeta_i U(r)
\eea
to make the linearised Einstein equations time-independent. Here, $E_{ai}$ are the external electric fields and $\zeta_i$ is thermal gradient which is defined as $\zeta_i=-(\nabla_i T/T)$. In the final expression for the each conserved currents, we will set $E_{1i}=E_{2i}=E_i$.  Since all the transports can be computed at the event horizon, we need to find the regularity condition at the horizon. We take the Eddington-Finkelstein coordinates $(v,r)$ where the background metric is regular at the horizon,
\bea
	ds^2=-Udt^2-2\sqrt{UV}dvdr+Wd\vec{x}^2
\eea
where $v=t+\int dr\sqrt{V/U}$. In this coordinates, the   metric perturabation  is given by 
\bea
	\delta g_{\mu\nu}dx^{\mu}dx^{\nu}=h_{tx}dvdx+\left( h_{rx}-\sqrt{\frac{V}{U}}h_{tx}\right)drdx.
\eea
To guarantee the regularity of the   metric with perturbation at the horizon, we demand the last term to vanish at the horizon so that
\bea 
	h_{ri}\sim\sqrt{\frac{V}{U}}h_{ti}.
\eea
The gauge fields can be reexpressed in the Eddington-Finkelstein coordinates to get the regularity condition at the event horizon:
\bea \label{horizon1}
	\delta B_{ai}\sim \tilde{b}_{ai}+E_{ai}v-E_{ai}\int dr\sqrt{\frac{V}{U}}.
\eea
Then, the full gauge field have the regular form of $\delta B_{ai}\sim E_{ai} v+\cdots$ in the Eddington-Finkelstein coordinates by demanding 
\bea
	\tilde{b}_{ai}'\sim\sqrt{\frac{V}{U}}E_{ai}.
\eea

We can define the radially conserved currents which is defined by
\bea\label{concurrent}
{J}_a^{\mu}&=&\sqrt{-g}Z_{a} G^{\mu r}_{(a)}+\frac{q_{\chi_a}}{4} \sum_{i}(\partial\chi_{i})^2\varepsilon^{\alpha\beta \mu r}G^{(a)}_{\alpha\beta},\nonumber\\
Q_i &=& \frac{U^2}{\sqrt{UV}}\left(\frac{h_{ti}}{U}\right)'-\sum_{a=1,2}b_aJ_{ai}
\eea
where the index $a=1,2$ denotes the two currents which are dual to the two gauge fields $B_a$ and $i=x,y$ is the index of directions. Since $J_a$ and $Q_i$ are the conserved quantities along the radial direction so that they can be evaluated at arbitary value. Hence, it is enough to compute at the horizon ~\cite{Donos:2014cya}. 

Finally, we can express the boundary current in terms of the external sources and transports coefficients:
\bea
J_{ai} &=& \sum_{bj}(\sigma_{ab})_{ij}E_{bj}+\sum_j(\alpha_a)_{ij}T \zeta_j\nonumber\\
Q_i&=&\sum_{aj}(\bar{\alpha}_a)_{ij} T E_{aj}+\sum_j\bar{\kappa}_{ij}T\zeta_j. \label{currentsource}
\eea
Before we express the transport coefficients explicitly, it will be useful to define following functions for the simple expression:
\bea
	\mathcal{F} &=& W Y \beta^2 + (Z_1+Z_2)H^2-\sum_{a=1,2}\frac{1}{Z_a}\left(Q_a \Theta_a H-\Theta_a^2 H^2\right)\nonumber\\
	\mathcal{G} &=& \sum_{a=1,2}\left(Q_a-\Theta_aH\right)
\eea	
where $\Theta_a=q_{\chi_a}\beta^2/W$.
One can define the total electric current as $J_i = \sum_a J_{ai}$ and identify the external electric field as $E_{ai}=E_i$. Then, each transport coefficient based on this total current and electric field is given by
\bea
	\sigma_{ij} &=& \frac{\partial J_i}{\partial E_j} = \sum_{ab} (\sigma_{ab})_{ij} \nonumber\\
			&=&\delta_{ij}\frac{Z\left(\mathcal{F}+\mathcal{G}^2/Z\right)\left(\mathcal{F}-Z H^2\right)}{\mathcal{F}^2+H^2\mathcal{G}^2}+\epsilon_{ij}\left(\Theta +\frac{ZH\mathcal{G}\left(2\mathcal{F}+\mathcal{G}^2/Z-ZH^2\right)}{\mathcal{F}^2+H^2\mathcal{G}^2}\right),\label{electriccond}\\
	\alpha_{ij}&=&\frac1T\frac{\partial J_i}{\partial \zeta_j} = \sum_a(\alpha_a)_{ij}\nonumber\\
		   &=& \delta_{ij}\frac{s \mathcal{G}\left(\mathcal{F}-ZH^2\right)}{\mathcal{F}^2+H^2\mathcal{G}^2}+\epsilon_{ij}\frac{sH\left(\mathcal{G}^2+Z\mathcal{F}\right)}{\mathcal{F}^2+H^2\mathcal{G}^2},\label{thermoelectric}\\
	\bar{\kappa}_{ij}&=&\delta_{ij}\frac{s^2T\mathcal{F}}{\mathcal{F}^2+H^2\mathcal{G}^2}+\epsilon_{ij}\frac{s^2TH\mathcal{G}}{\mathcal{F}^2+H^2\mathcal{G}^2}
\eea
where $Z=Z_1+Z_2$ and $\Theta=\Theta_1+\Theta_2$. Notice that $\bar{\alpha}_{ij}=\alpha_{ij}$.
The resisitivity is defined as the inverse of the conductivity matrix:
\bea
	\rho_{ii}&=&\frac{\sigma_{ii}}{\sigma_{ii}^2+\sigma_{ij}^2}=\mathcal{R}_{ii}/\mathcal{D},\nonumber\\
	\rho_{ij}&=&\frac{\sigma_{ij}}{\sigma_{ii}^2+\sigma_{ij}^2}=\mathcal{R}_{ij}/\mathcal{D}
\eea
where
\bea
	\mathcal{R}_{ii}&=&(\mathcal{G}^2+Z\mathcal{F})(\mathcal{F}-ZH^2),\nonumber\\
	\mathcal{R}_{ij}&=&(\mathcal{F}^2+H^2\mathcal{G}^2)\Theta+H\mathcal{G}(\mathcal{G}^2+2Z\mathcal{F}-Z^2H^2),\nonumber\\
	\mathcal{D}&=&(Z\mathcal{F}+\mathcal{G}^2)	^2+(\mathcal{F}^2+\mathcal{G}^2H^2)\Theta^2+H(Z^2H-2\mathcal{G}\Theta)(Z^2H^2-2Z\mathcal{F}-\mathcal{G}^2).
\eea
The thermal conductivity $\kappa$ is defined by the response of the temperature gradient $T\zeta_i$ to the heat current $Q_i$ in the absense of the electric currents $J_{ai}$. Setting $J_{ai}=0$ in (\ref{currentsource}), we can write $E_{bj}$ in terms of $\zeta_j$ to substituting to the expression of the heat current in (\ref{currentsource}). Then, we can get
\bea
	\kappa=\bar{\kappa}-T\left(\bar{\alpha}_1(\alpha_1\sigma_{22}-\alpha_2\delta)+ \bar{\alpha}_2(\alpha_2\sigma_{11}-\alpha_1\delta)\right)(\sigma_{11}\sigma_{22}-\delta^2)^{-1}
\eea
where $\delta=\sigma_{12}=\sigma_{21}$. Notice that this expression is very similar to that in ~\cite{Seo:2016vks}, but it is $2\times 2$ matrices multiplication which is different from the simple scalar multiplication in ~\cite{Seo:2016vks}.

The Seebeck coefficient $S$ and the Nernst signal $N$ are given by
\bea
	S&=&-(\Sigma^{-1}\cdot\mathcal{A})_{xx},\nonumber\\
	N&=&-(\Sigma^{-1}\cdot\mathcal{A})_{yx}
\eea
where
\bea
	\Sigma=
	\begin{pmatrix}
		\sigma_{11} & \delta\\
		\delta & \sigma_{22}
	\end{pmatrix},\qquad 
	\mathcal{A}=
	\begin{pmatrix}
		\alpha_1\\
		\alpha_2
	\end{pmatrix}
\eea
Here, $\Sigma$ and $\mathcal{A}$ are $4\times4$ and $4\times2$ matrices respectively.

 As we discussed in the Introduction, we need the two currents as the independently conserved currents which is identified with $J_e$ and $J_h$. The total electric current $\vec{J}$ and the total number current $\vec{J}_n$ are defined by $\vec{J}=\sum_{a} \vec{J}_a\equiv\vec{J}_e+\vec{J}_h$ and $\vec{J}_n=\vec{J}_e-\vec{J}_h$ respectively. Their corresponding densities are related by $Q_1=q_e n_1$ and $Q_2=-q_en_2$ with a charge of electron $q_e=-1$. The total electric charge density and the total number density are defined by $Q=Q_1+Q_2$ and $Q_n=-Q_1+Q_2$ and we can connect each density with the proportional constant $g_n$ where $Q_n=g_nQ$. We have the simple expressions for the case in the absence of the external magnetic field as following~\cite{Song2020}:
 \bea
 	\sigma_{xx}= Z\left(1+\frac{Q^2}{Q_0^2}\right),\qquad \kappa_{xx}=\frac{\bar{\kappa}_{xx}}{1+(1+g_n^2)(\frac{Q^2}{Q_0^2})}
 \eea
 where $Q_0^2=WYZ\beta^2$.
 
\begin{figure}[h]
\begin{center}
\subfigure[]{\includegraphics[angle=0,width=0.32\textwidth]{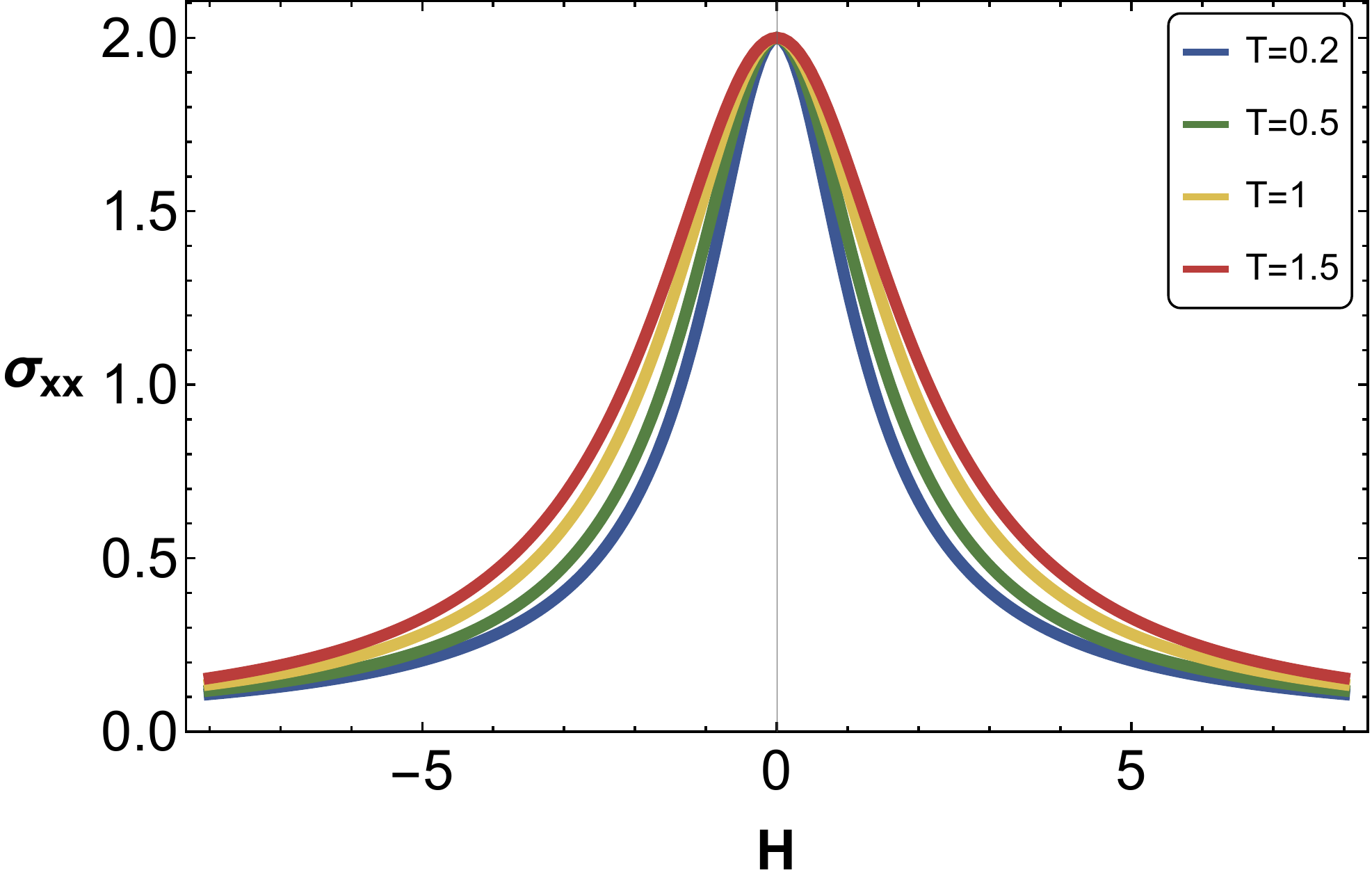}}
\subfigure[]{\includegraphics[angle=0,width=0.32\textwidth]{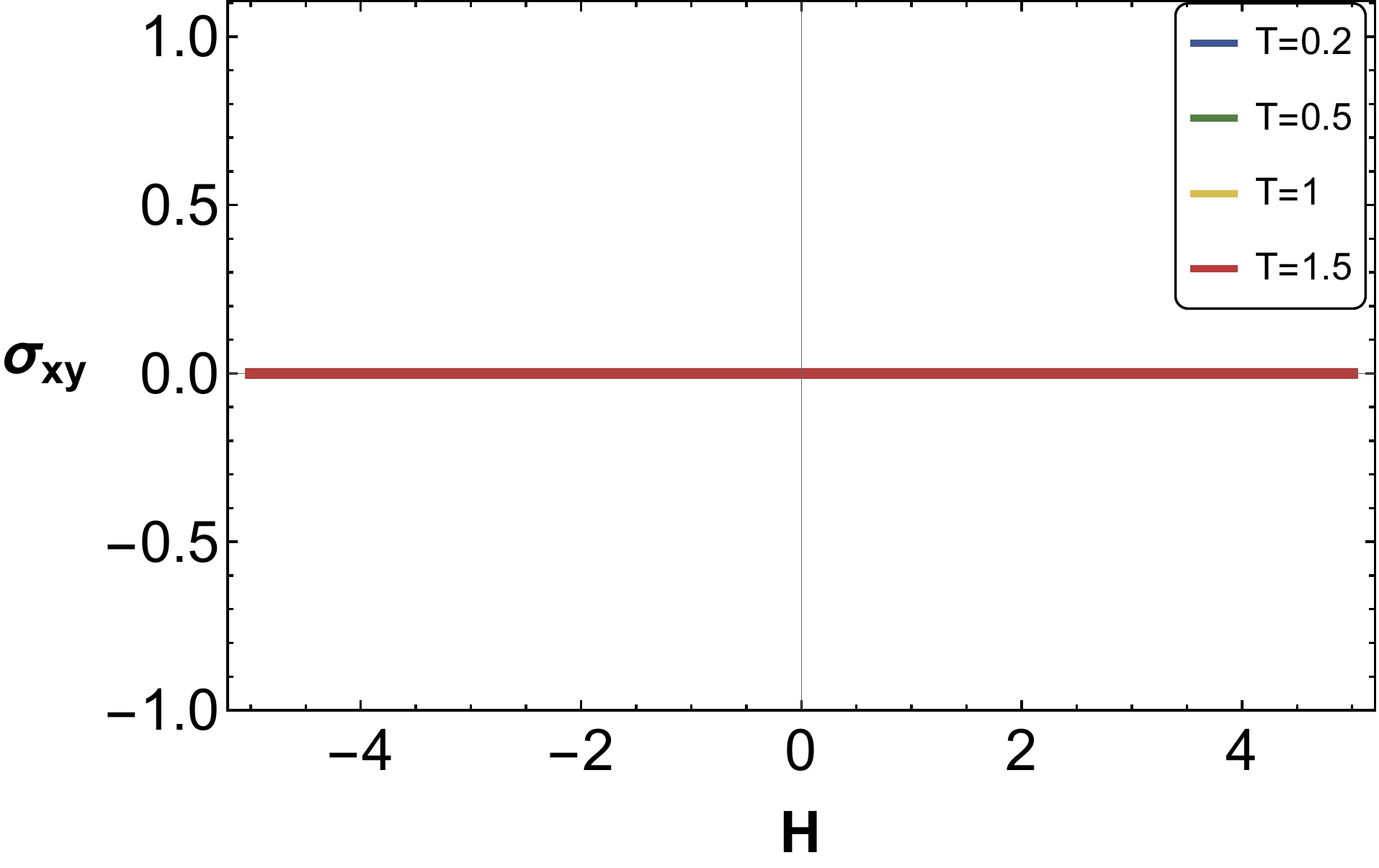}}
\subfigure[]{\includegraphics[angle=0,width=0.32\textwidth]{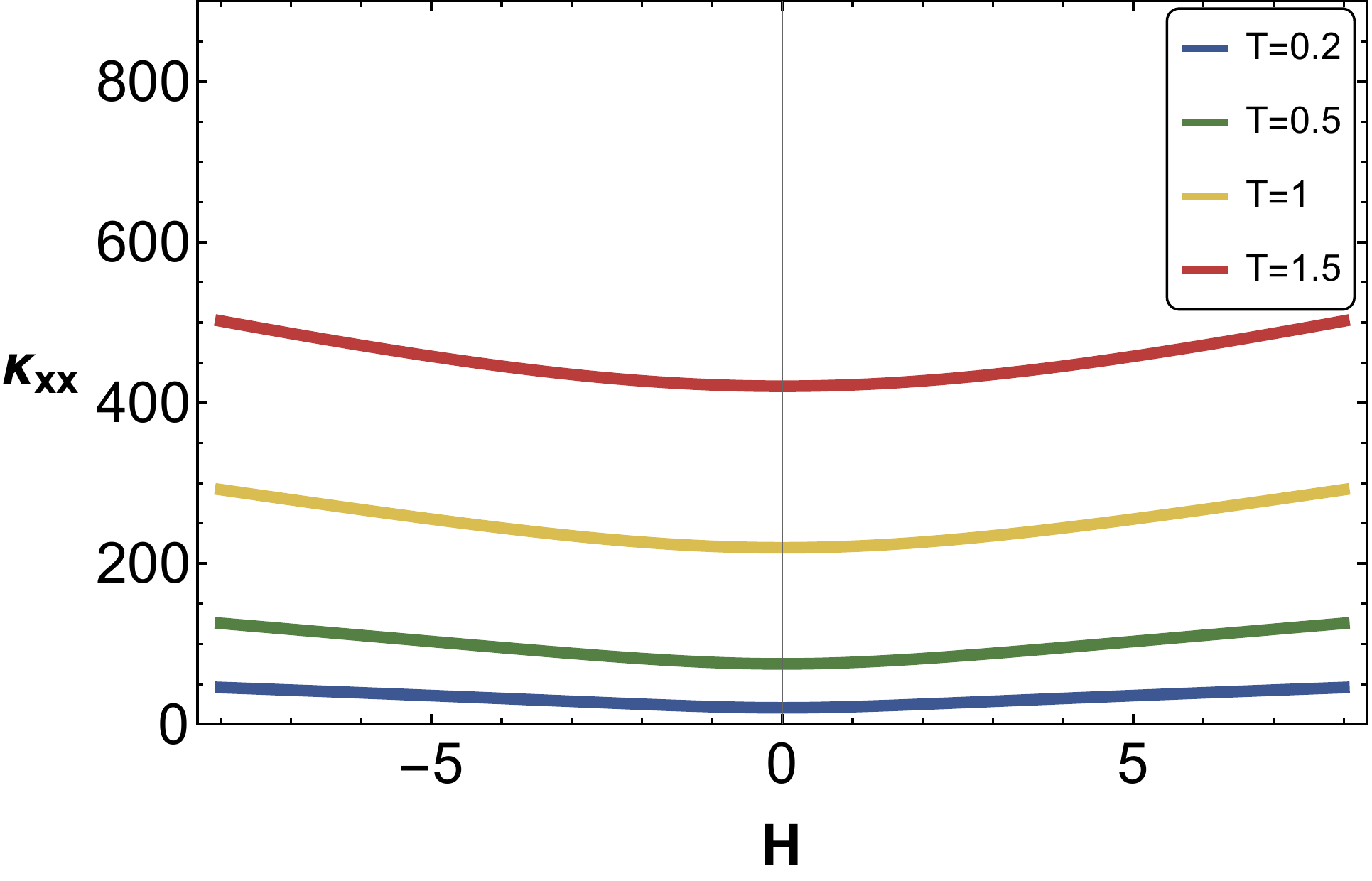}}
\caption{Magnetotransport for $z=3/2$, $\theta=1$ without the magnetic impurities. We choose the parameters as $\bar{Z}_1=\bar{Z}_2=1$, $q_{\chi_1}=0$, $q_{\chi_2}=0$, $g_n=3$ and $\beta=1.5$.
   } \label{fig1} 
\end{center}
\end{figure}

\begin{figure}[h]
\begin{center}
\subfigure[]{\includegraphics[angle=0,width=0.32\textwidth]{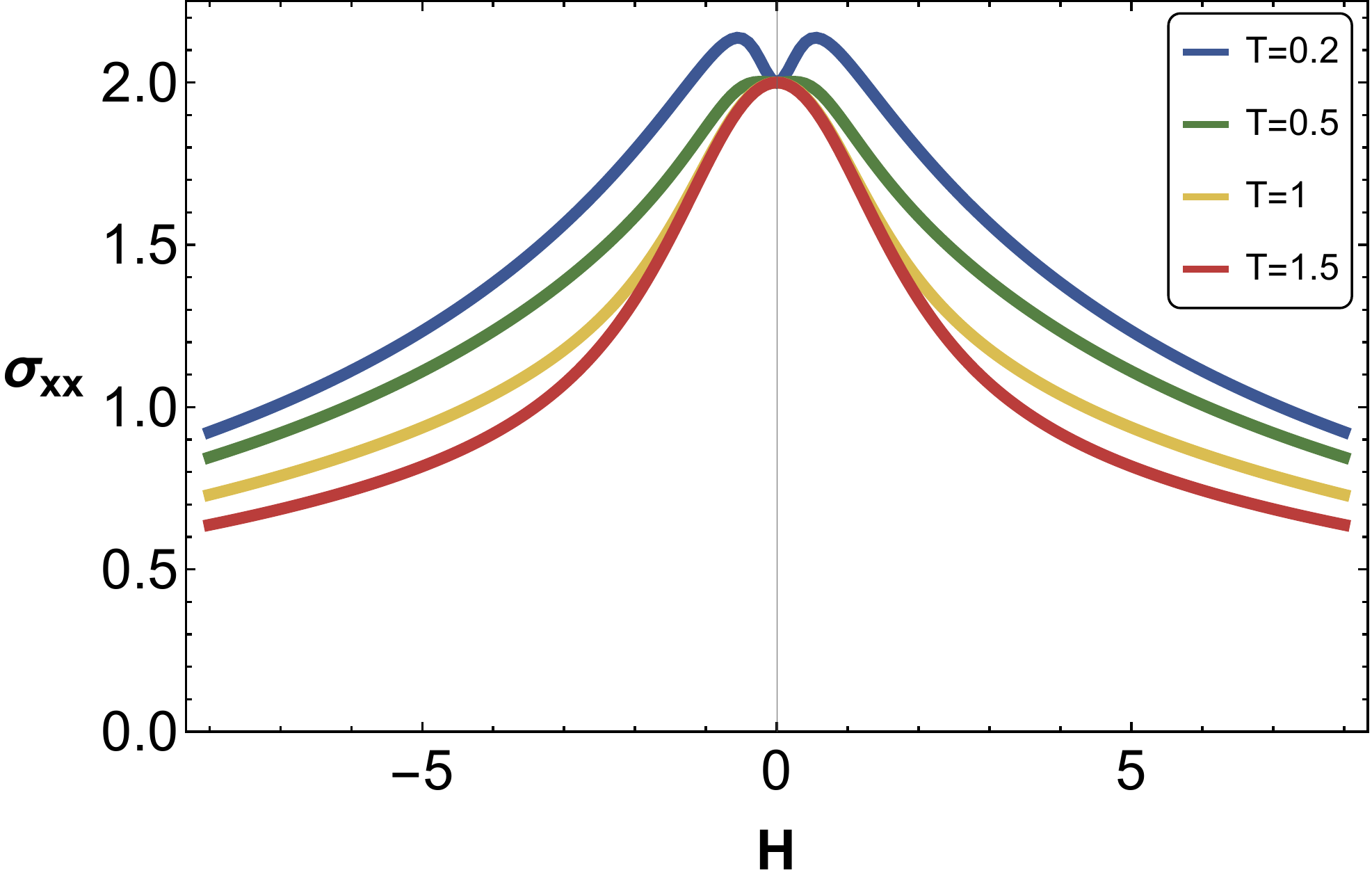}}
\subfigure[]{\includegraphics[angle=0,width=0.32\textwidth]{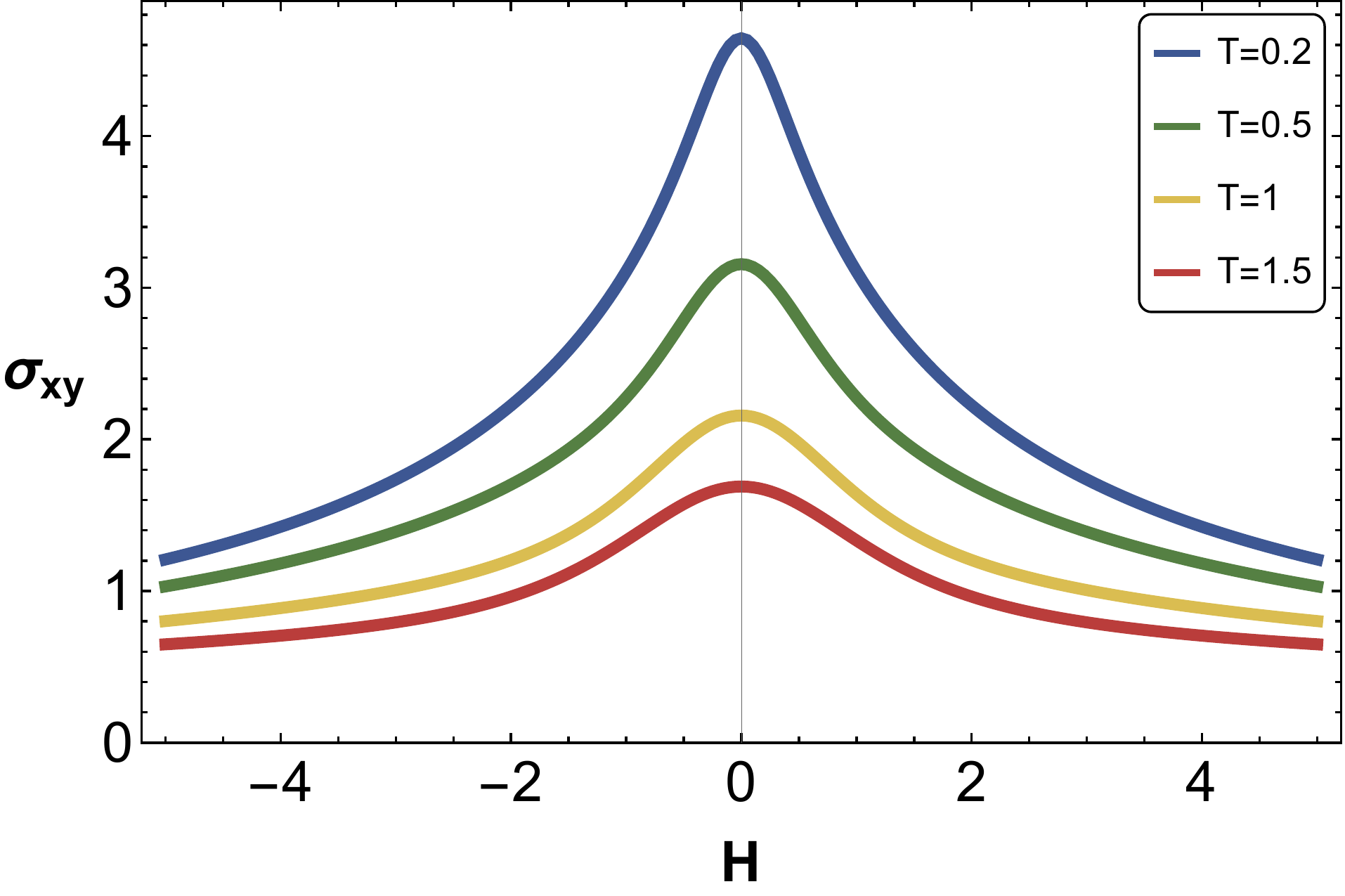}}
\subfigure[]{\includegraphics[angle=0,width=0.32\textwidth]{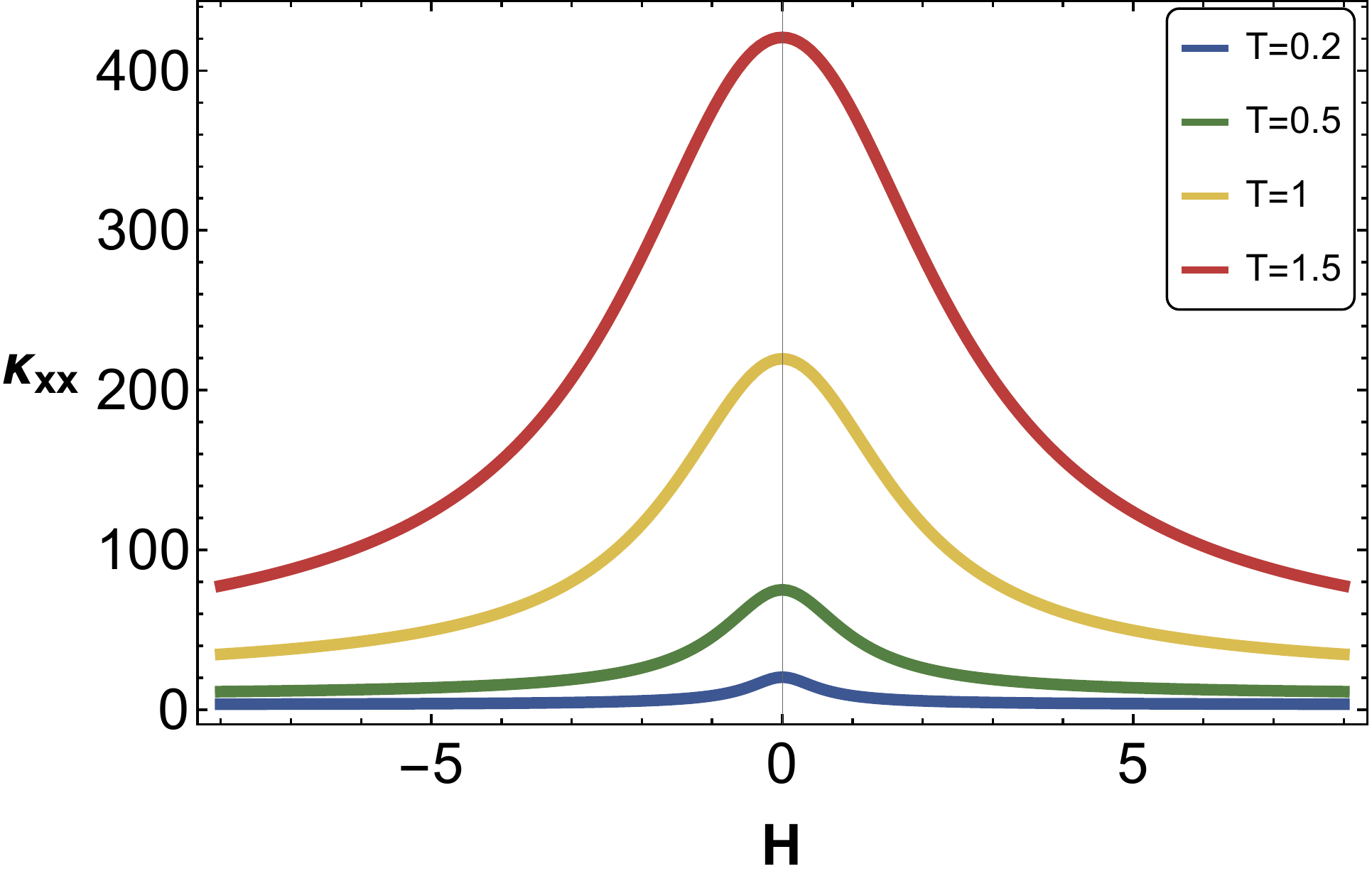}}
\caption{Magnetotransport for $z=3/2$, $\theta=1$ with the magnetic impurities. We choose the parameters as $\bar{Z}_1=\bar{Z}_2=1$, $q_{\chi_1}=1$, $q_{\chi_2}=2$, $g_n=3$ and $\beta=1.5$.
   } \label{fig2} 
\end{center}
\end{figure}

In figures \ref{fig1}, \ref{fig2}, we show the typical behaviors of each magnetotransport when $z=1.5$, $\theta=1$ and $Q_1=Q_2=0$ where we are interested. Notice that $\kappa_{xy}=0$ when there's no conserved electric charge.
As one can see from Figure \ref{fig1} and \ref{fig2},  there is no significant difference between a single current model and two currents model in the qualitative sense compared to the results in ~\cite{Ge2020}. But, as in ~\cite{Seo:2016vks, Song2020},  the two currents model can give the physical implication if there is experimental data to compare with this model.  Unfortunately, we don't have any relevant experiments to be conducted so that we leave our results as a qualitative prediction for experiment for the graphene with magnetic doping.

\begin{figure}[h]
\begin{center}
	\subfigure[$g_n=0$]{\includegraphics[angle=0,width=0.42\textwidth]{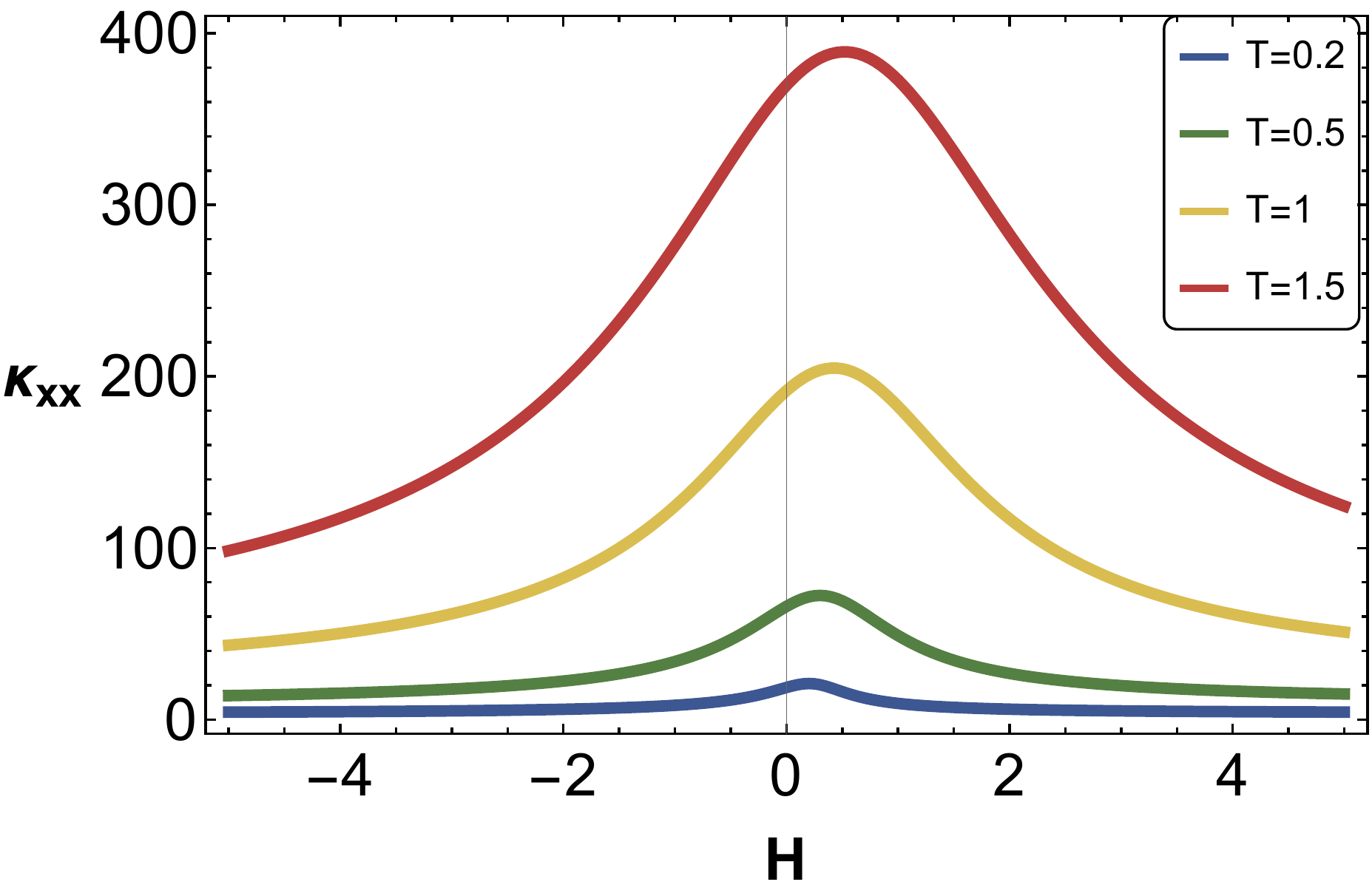}}
	\subfigure[$g_n=1.5$]{\includegraphics[angle=0,width=0.42\textwidth]{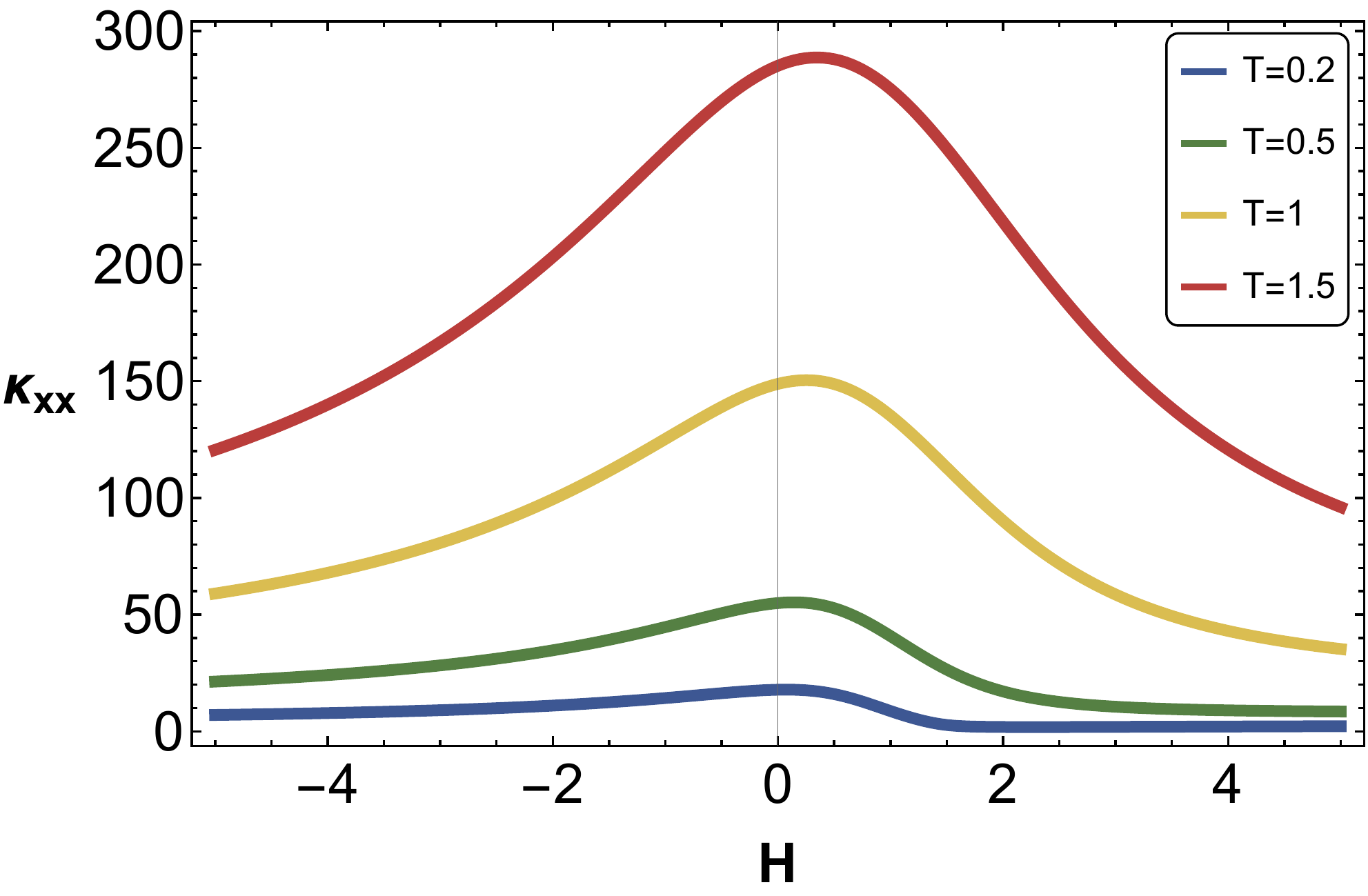}}
\caption{Magnetotransport for $z=3/2$, $\theta=1$ with the magnetic impurities. We choose the parameters as $\bar{Z}_1=\bar{Z}_2=1$, $q_{\chi_1}=1$, $q_{\chi_2}=2$, $Q=2$ and $\beta=1.5$.
   } \label{fig3} 
\end{center} 
\end{figure}

Finally, we set $Q\neq0$ to see the effect of $g_n$ which correponds to two current effect. (See Figure \ref{fig3})
Due to the presence of the charge, we have the assymmtric feature in $\kappa_{xx}$ and this assymetry is enhanced by $g_n$. For the larger $g_n$, we have the non-analytic behavior which comes from (\ref{r0Trelation}). But, we would like to leave the analysis for this non-analyticity to the future work.
\section{Conclusion} 
In this paper, we investigated the two currents model with magnetic doping in the presence of the magnetic field, which is based on ~\cite{Seo:2016vks,Song2020,Seo:2017oyh,Ge2020}. From this model, we calculated all transport coefficients. Although we do not expect   qualitative difference between a single current and two currents model, the presence of two current are definitely necessary to describe the quantitative data fitting for the material which involve two independent electron system that are very weakly coupled, 
 like graphene   ~\cite{Seo:2016vks,Song2020} or other multi-valley systems or multi-layer systems  which will be  studied in a future.  


\acknowledgments
 This  work is supported by Mid-career Researcher Program through the National Research Foundation of Korea grant No. NRF-2016R1A2B3007687.

\bibliographystyle{JHEP}
\bibliography{Refs_2J.bib}

\end{document}